# Loss Dependence on Geometry and Applied Power in Superconducting Coplanar Resonators

Moe S. Khalil, F. C. Wellstood, and Kevin D. Osborn

*Abstract*— The loss in superconducting microwave resonators at low-photon numbers and low temperatures is not well understood but has implications for achievable coherence times in superconducting qubits. We have fabricated single-layer resonators with a high quality factor by patterning a superconducting aluminum film on a sapphire substrate. Four resonator geometries were studied with resonant frequencies ranging from 5 to 7 GHz: a quasi-lumped element resonator, a coplanar strip waveguide resonator, and two hybrid designs that contain both a coplanar strip and a quasi-lumped element. Transmitted power measurements were taken at 30 mK as a function of frequency and probe power. We find that the resonator loss, expressed as the inverse of the internal quality factor, decreases slowly over four decades of photon number in a manner not merely explained by loss from a conventional uniform spatial distribution of two-level systems in an oxide layer on the superconducting surfaces of the resonator.

## I. INTRODUCTION

THE growing interest in superconducting quantum computing and microwave kinetic inductance detectors (MKIDS) has motivated the extensive study of superconducting resonators at milli-Kelvin temperatures in the limit of low photon numbers [1]-[9]. Thin-film resonators operating in this limit are potentially useful for photon detection [10] and as components for storing and transferring information between qubits [11]-[13]. It has been found that in the single-photon regime, some resonators are limited by amorphous dielectric loss due to two-level system (TLS) defects and that this type of loss can be an important source of decoherence in superconducting phase qubits [14]. However, the loss in coplanar resonators on crystalline dielectrics such as sapphire and crystalline silicon is less clear and is more difficult to locate due to non uniform field distributions. Many such resonators have been studied and the phase noise [1], [2], and loss [7], [9], are often modeled as a surface distribution of TLSs.

Most previous research in this area has focused on quasi-one-dimensional cavity resonators, such as coplanar waveguide transmission line resonators [11]-[13]. Lumped element devices are less popular but are sometimes used in quantum information, like in qubits [15], [16], and in Josephson junction resonators [17]. There is also growing interest in coupling quasi-lumped element resonators to qubits [18] because the lack of harmonic modes reduces loss from the Purcell effect [19]. When coupling to a qubit, it's been found that the symmetry of the qubit must be considered, since the type of coupling to it may affect coherence [20].

Here we present measurements on the internal quality factor $Q_i$, of four distinct coplanar superconducting resonators between 5 and 7 GHz, which include both quasi-lumped and quasi-one-dimensional cavity transmission-line resonators. Their symmetric shape induces inductive, rather than capacitive, coupling to them.

## II. RESONATOR DESIGN AND FABRICATION

All resonators were fabricated with 100 nm thick sputtered aluminum films on c-plane sapphire wafers. The aluminum was patterned with positive photoresist and wet etched in a bath composed mainly of phosphoric and nitric acid.

One of the resonators is a quasi-lumped element resonator, composed of a meandering quasi-lumped inductor (QLL) and an interdigital quasi-lumped capacitor (QLC), shown in Fig. 1(a), with an inductance and a capacitance of approximately 2 nH and 0.3 pF respectively. Another is a 4.5mm long shorted λ/4 length coplanar strip (CPS) resonator, shown in Fig. 1(b). Unlike in the quasi-lumped (QL) resonator the electric field in the CPS is distributed across the length of the resonator rather than being confined to a single element. The two others have both a CPS element and a QL element, Fig. 1(c, d).

The four resonators were embedded in the ground plane of the same 50 Ω coplanar waveguide (CPW) and were coupled inductively (rather than capacitively) to that waveguide (see Fig. 2). Since these resonators primarily modify the transmission near resonance, they form a four-notch band-block transmission filter.

The point connecting the two nominally symmetric halves of the resonator (circle 1 in Fig. 1) is at the current anti-node and voltage node, and the sides of the resonators far from the coplanar waveguide (circle 2 in Fig. 1) are at the voltage anti-nodes. The fundamental resonance frequency is antisymmetric in all four resonators. The effective capacitive coupling at this mode is weak, due to the design of the structures.

M. S. Khalil is with the Laboratory for Physical Sciences, College Park, MD 20740 USA and with the Center for Nanophysics and Advanced Materials, Department of Physics, University of Maryland College Park, MD 20740 USA (e-mail: moe@lps.umd.edu).

Kevin D. Osborn is with the Laboratory for Physical Sciences, College Park, MD 20740 USA (e-mail: osborn@lps.umd.edu).

F. C. Wellstood is with the Joint Quantum Institute and the Center for Nanophysics and Advanced Materials, Department of Physics, University of Maryland College Park, MD 20740 USA (e-mail: well@squid.umd.edu).

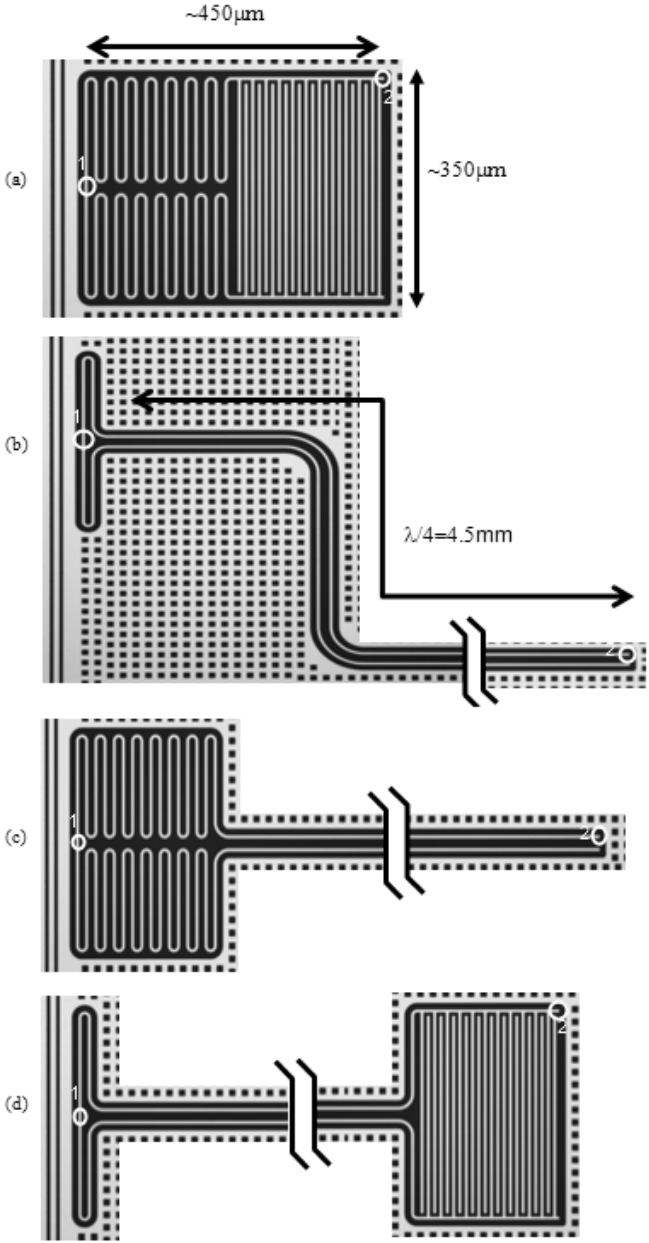

Figure 1. Optical images of four resonators measured (gray/white is aluminum metal and black is sapphire substrate). Circle 1 labels the current anti-node and voltage node of resonators and Circle 2 labels one of the two current nodes and voltage anti-nodes. (a) Quasi-lumped (QL) resonator at 5.46 GHz. (b) Coplanar strip (CPS) resonator at 6.44 GHz. (c) Quasi-lumped inductor with a CPS (QLL-CPS) resonator at 5.76 GHz. (d) Quasi-lumped capacitor with a coplanar strip (QLC-CPS) resonator at 6.01 GHz.

This can be understood from the nearly symmetric shape of these structures, which implies the capacitances to either side of the resonators are approximately equal. As a result, the capacitance cannot couple to antisymmetric modes. Since the lowest frequency mode is antisymmetric we expect only inductive coupling between the resonators and the CPW. An example of this is shown in Fig. 2(f), where a capacitive network couples voltage V to a resonator. With a symmetric design, C1=C2, C3=C4, C5=C6, C7=C8, the coupling cannot excite antisymmetric modes.

The structure of the resonance was confirmed with an EM simulator. It should be pointed out that while the symmetry of the resonances is the same for all four resonators, the dimension of resonances is not. The CPS resonator is a quasi-one-dimensional cavity, while the QL resonator acts like a quasi-zero-dimensional cavity. This is clear when simulating the full spectrum of these structures, because for the CPS resonator higher-order harmonics occur every half wavelength but for the QL resonator no higher order modes occur until frequencies above 30 GHz.

Near the resonance frequency each of the four resonators can be represented by an equivalent lumped LC circuit. The Norton equivalent circuit is shown in Fig. 2(g). The internal loss components such as TLSs discussed earlier, radiation, or metal loss can be represented (for the purposes of this paper) by the resistor R and act as a loss component lowering internal quality factor $Q_i$. The coupling of the resonator to the CPW can be represented by a resistor $R_T$ and acts as a loss component with coupling quality factor $Q_e$. These quality factors add reciprocally to give the total measured quality factor $Q_T$, $1/Q_T = 1/Q_i + 1/Q_e$.

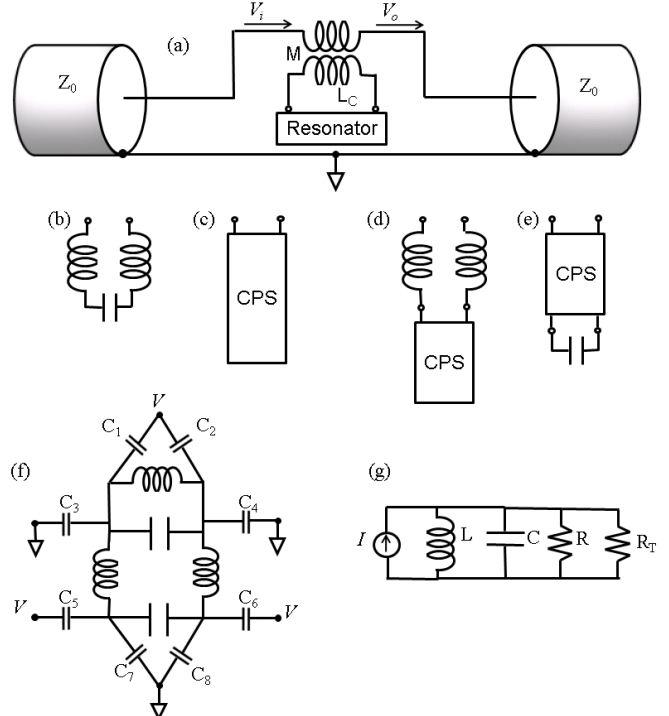

Figure 2. (a) Schematic drawing of an arbitrary resonator inductively coupled to a coplanar waveguide. (b-e) Schematic drawing of each of the four types of resonators. (b) QL. (c) CPS. (d) QLL-CPS. (e) QLC-CPS. (f) Schematic of the capacitive coupling to these resonators. Due to the symmetry of these resonators $C_1=C_2$, $C_3=C_4$, $C_5=C_6$, and $C_7=C_8$, capacitive coupling cannot excite an antisymmetric resonance. As a result, only inductive coupling remains. (g) Norton equivalent of a resonator coupled to the waveguide.

III. EXPERIMENTAL SETUP AND MEASUREMENT

Measurements were performed at 30 mK on the mixing chamber of a dilution refrigerator unless otherwise specified. The input transmission line had 20 dB of resistive attenuation at both the 1 K and 30 mK stages. The output transmission line had a circulator with 18 dB of isolation on each of the same two temperature stages. A low-noise Caltech HEMT amplifier was placed at 4 K on the output line.

The output signal can be written as,

$$V_0/V_i = 1 - \frac{Q_T/Q_e}{1 + i2Q_T(\omega - \omega_0)/\omega_0}. \quad (1)$$

Where $\omega$ is the measurement frequency, $\omega_0$ is the resonance frequency, $Q_T$ is the total quality factor of the resonator and $Q_e$ is the external quality factor. In general $Q_e$ can have an imaginary component to account for a small additional inductance at the CPW near the resonator and possible impedance mismatches resulting in standing waves in the measurement. Eq. (1) is used to fit our $S_{21}$ measured data, shown in Fig. 3, from which we can extract all three relevant quality factors and the resonance frequency, $\omega_0$. $Q_e$ is a function of the coupling ($1/Q_e \sim M^2$) and should be power independent, which serves as a good check for our fits. Indeed, we find $Q_e$ to be power independent, even over a broad range of input powers. Note that the imaginary component of $Q_e$ appears graphically as an asymmetry in the measured resonance Lorentzian, or in a plot of the real versus imaginary components of the transmission as a rotation of the resonance circle around the off resonance point.

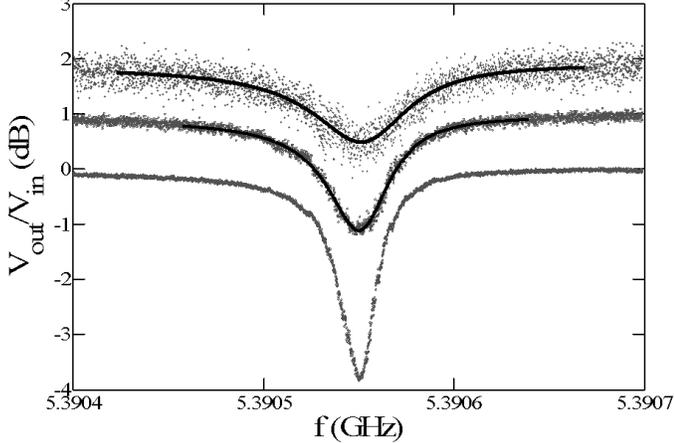

Figure 3. Raw transmission measurements with fits. The three Lorentzians are three different drive powers. Note, for data in this paper we only fit to the amplitude of the output voltage. The phase can be extracted from (1) and can also be used for fitting. Both amplitude and phase data can be combined into an in-phase vs. quadrature plot where the resonance produces a circle such that the point on the circle closest to the origin is the resonance frequency.

## IV. THEORY OF LOSSES

It has been found that in most amorphous bulk dielectrics such as $SiN_x$ and $SiO_x$ the loss can be attributed to TLS defects [14]. For loss from tunneling TLSs with a single relaxation time $T_1$ and coherence time $T_2$, one expects the loss tangent to obey [21],

$$\tan\delta = \tan\delta_0 \frac{\tanh\left(\frac{\hbar\omega}{2kT}\right)}{\sqrt{1 + \left(E/E_C\right)^2}}, \quad (2)$$

where $\tan\delta_0$ is the low-power intrinsic loss tangent dependent on the frequency distribution and density of the TLSs, $E_C$ is the critical field dependent on the $T_1$ or $T_2$ of the TLSs, and $E$ is the applied electric field at the location of the TLSs.

The inverse of the internal quality factor depends on the weighted distribution of the loss tangent:

$$\frac{1}{Q_i} = \frac{P_{loss}}{\omega U} = \frac{\int_{\substack{Lossy\\Material}} \tan\delta(\varepsilon(\vec{r}))(\vec{E}(\vec{r}))^2 d^3r}{\int_{\substack{All\\Space}} \varepsilon(\vec{r})(\vec{E}(\vec{r}))^2 d^3r}. \quad (3)$$

We have evaluated (3) for TLSs on the metal sidewalls, the metal top surface and the metal-dielectric interface for both the CPS and the QLC geometries. We do this by first finding the electric field as a function of position for these structures using the finite element EM simulator COMSOL. Then we use a uniform spatial distribution of TLS from (2) in the evaluation of (3). We find that for all resonant structures even for a surface distribution of TLSs, no matter the surface, $1/Q_i$ never decreases slower than $1/V^{0.8}$ for more than a decade in voltage.

## V. RESULTS

The resonators were measured in three cool-downs with three different chips from two different wafers. Fig. 4 shows results for two of the resonators on two chips from the same wafer. We compare the CPS resonators in Fig. 2(b) and the QLL-CPS resonators in Fig. 2(d). Fig. 5 shows the measurement for a third chip from a different wafer. The loss is within a factor of 2 for all of the resonators, with the best resonator (a CPS resonator) having an internal quality factor of 200,000 at single photon numbers (dashed lines in Fig. 4 and Fig. 5) and 1.1 million at $10^6$ photon numbers (note that the number of photons stored is proportional to $V^2$).

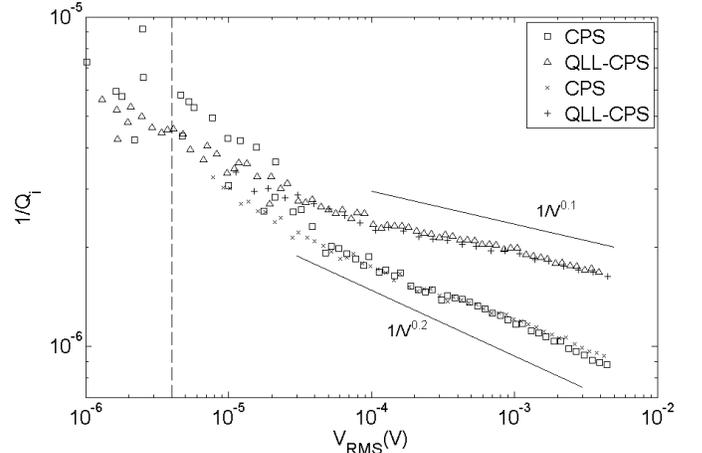

Figure 4. Comparing two resonators, the CPS and the QLL-CPS hybrid, from two different chips from the same wafer. The loss in the resonators appears to be consistent from chip to chip, with the CPS showing slightly greater power dependence. Loss can be approximated at high powers by a power law of about $1/V^{0.1}$ to $1/V^{0.2}$ depending on the resonator. The dashed line indicates a single photon of excitation (4 x $10^{-6}$V).

Data in Fig. 4 shows lower loss in the CPS resonator than the QLL-CPS hybrid resonator for two chips on the same wafer. However, after inspecting Fig. 5, showing data for all four resonators from two different wafers, it is clear that the relative loss is not consistent from wafer to wafer. While the loss in all of the resonators is very low, the loss varies between wafers (but apparently not between chips on the same wafer) by an amount greater than our experimental precision. This indicates that there is an uncontrolled change in the loss versus power data between wafers, which could for example be explained by variations in the fabrication process from wafer to wafer.

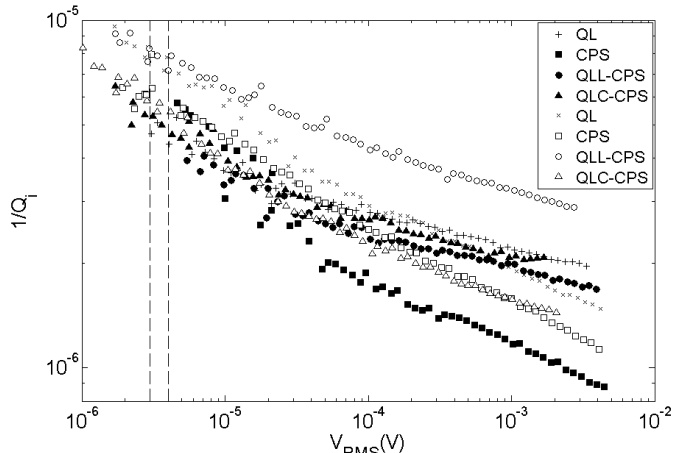

Figure 5. Comparing all four resonators from two different chips and two different wafers. The magnitude of the loss for individual resonators seems to change and they also switch their respective hierarchy of losses. Dashed lines indicate the range of single photon excitation (3 x $10^{-6}$ V or resonators with a QLC and 4 x $10^{-6}$ V for resonators without a QLC).

Despite the variation from sample to sample, the power dependent loss tangent at the highest powers for all four resonators can be approximated by a power law between $1/V^{0.1}$ and $1/V^{0.2}$ over several decades in photon number. Two lines representing $1/V^{0.1}$ and $1/V^{0.2}$ power laws are drawn in Fig. 4. This is a much weaker power dependence than one would find in a parallel plate capacitor $(1/V)$ [14]. It is also shallower than anything that can be produced assuming a uniform surface and frequency distribution of TLSs with a single critical electric field. As noted above, the shallowest power dependence we obtain for a spatial distribution of TLS on a surface is greater than $1/V^{0.8}$.

A notable feature of the measured loss (not shown) is that it does not depend on temperature between 30 and 200 mK. Above 200 mK, loss from thermal quasiparticles in the superconducting films limits the quality, and one finds reasonable agreement with Mattis-Bardeen theory [22]. The behavior below 200 mK is puzzling because the predicted temperature change in (2) should be 40% over this range, well within our experimental precision.

In conclusion, we measured four inductively coupled superconducting coplanar resonators containing both quasi-lumped elements and transmission line elements. We found all resonators had high quality factors with no resolvable differences in the loss of the different types of resonators. This implies that all four resonator designs are appropriate for quantum computing circuits, and there does not seem to be a large advantage for one type of resonator design over another if only considering the fundamental mode. However, since the quasi-lumped element resonator does not have harmonic cavity modes, it may still provide an advantage when measuring the lifetime of a coupled qubit [18], [19]. We also found that the power dependence of the loss in these resonators is not consistent with a conventional surface distribution of identical TLSs.

ACKNOWLEDGMENTS

The authors acknowledge helpful discussions with M. Stoutimore, S. Gladchenko, B. Sarabi, C. Lobb, B. Palmer, Z. Kim, V. Zaretskey, and H. Paik.